\documentclass[twocolumn,showpacs,showkeys,amsmath,amssymb,floatfix,superscriptaddress]{revtex4}

\usepackage{graphicx}
\usepackage{dcolumn}

\begin{document}

\title{Fast Diffusion Mechanism of Silicon Tri-interstitial Defects}
\author{Yaojun A. Du}
\affiliation{Department of Physics, Ohio State University, Columbus, OH, USA}
\author{Stephen A. Barr}
\affiliation{Department of Materials Science and Engineering, 
University of Illinois at Urbana-Champaign, Urbana, Illinois, USA}
\author{Kaden R. A. Hazzard}
\affiliation{Department of Physics, Cornell University, Ithaca, NY, USA}
\author{Thomas J. Lenosky}
\affiliation{Department of Physics, Ohio State University, Columbus, OH, USA}
\author{Richard G. Hennig}
\affiliation{Department of Physics, Ohio State University, Columbus, OH, USA}
\author{John W. Wilkins}
\affiliation{Department of Physics, Ohio State University, Columbus, OH, USA}

\date{\today}

\begin{abstract}
  Molecular dynamics combined with the nudged elastic band method
  reveals the microscopic self-diffusion process of compact silicon
  tri-interstitials.  Tight-binding molecular dynamics paired with
  \textit{ab initio} density functional calculations speed the
  identification of diffusion mechanisms. The diffusion pathway can be
  visualized as a five defect-atom object both translating and
  rotating in a screw-like motion along $\langle 111 \rangle$
  directions.  Density functional theory yields a diffusion constant
  of $ 4 \times 10^{-5}~\exp (- 0.49~{\rm eV} / k_{B} T)~ {\rm cm^2/s}
  $.  The low-diffusion barrier of the compact tri-interstitial may be
  important in the growth of ion-implantation-induced extended
  interstitial defects.
\end{abstract}


 \maketitle

Following high-energy ion implantation, strongly supersaturated
silicon self-interstitials can agglomerate to form macroscopic planar
interstitial structures, $\{311\}$ defects~\cite{salisbury79pm}.
High-temperature annealing is necessary to remove lattice damage
following ion implantation.  However during the annealing process the
$\{311\}$ defects comprise a large reservoir of
interstitials~\cite{takeda91jjap,kohyama92prb,jnkim97prb}, which are
released upon annealing and hence drive boron transient enhanced
diffusion~\cite{eaglesham94apl,stolk97jap,zhang95apl}, an undesirable
process which causes spatial broadening of boron concentration
profiles.  On the other hand, following low-energy implantation,
significant boron TED is still observed, even though no visible
$\{311\}$ defects are developed~\cite{zhang95apl}.

\begin{figure}
\includegraphics[width=8cm]{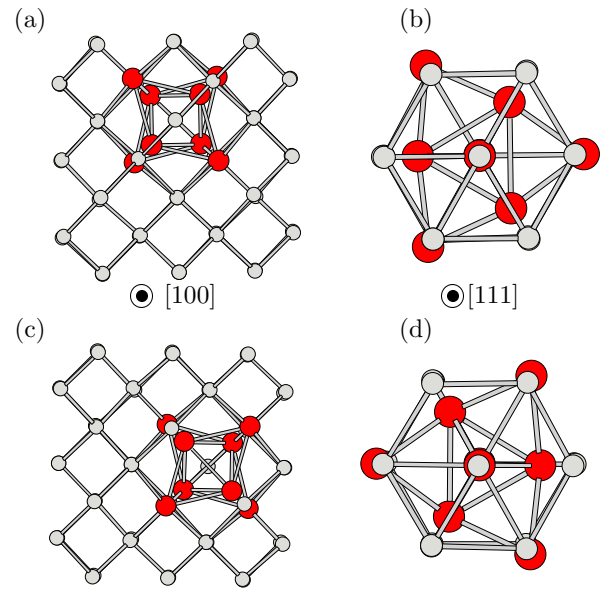}
\caption{\label{figure_I3b} Diffusion of $I_{3}^{b}$  to the neighboring site
  along the $\langle 111 \rangle $ direction. The initial
  configurations viewed from [110] and [111] are shown in (a) and (b),
  respectively.  The final configurations viewed from [110] and [111]
  are shown in (c) and (d), respectively. Defect atoms are shaded.
  All of the above structures are relaxed using density functional
  theory.  }
\end{figure}

It is crucial to understand the diffusion of various point defects in
order to have a quantitative understanding of boron TED.  Measurements
of the diffusion rate of defects in silicon have been reported in
experiments~\cite{tan85apa, bracht95prb, zimdat, boitdat, bronnerdat,
  wijdat, partyka}.  Current experimental techniques cannot cleanly
resolve the diffusion rates of more complex defect species, and due to
the atomic-scale size of point defects, defect diffusion pathways
cannot be resolved at all \cite{benton97jap}.  Thus, numerical
simulations provide a unique way to study the
technologically-important dynamics of point defects in crystalline
silicon.

Previously, such simulations have studied numerous defect species
using both classical and quantum
Hamiltonians~\cite{byy83prl,jnkim99prl, estreicher01prl,
  estreicher01pcab, birner01ssc}.  Using nudged elastic band methods
(NEB)~\cite{neb} within density functional theory (DFT) code, Lopez
\textit{et al.}  compute an activation energy of 0.28 eV for a neutral
single interstitial diffusing along the {\it X-T-X}
path~\cite{lopez04}.  Kim {\it et al.}  estimates the self-diffusion
barrier of a particular di-interstitial performing a reorientation to
be 0.5 eV~\cite{jnkim99prl}. Recent DFT-NEB calculations done in our
group confirm this pathway but give a somewhat lower barrier of 0.3
eV~\cite{pending}.

Cogoni \textit{et al.}~\cite{cogoniprb05} use temperature-accelerated
molecular-dynamics~\cite{TAD} with the Kwon {\it et al.}~\cite{kwon}
tight-binding (TB) potential to systematically study the diffusion of
low-lying single-, di- and tri-interstitials.  In particular they find
all are fast diffusers with diffusion barriers of 0.94, 0.89 and
1.71~eV, respectively.  We extend that work both with a more accurate
tight-binding potential and with density-functional theory to further
refine the diffusion pathway and the diffusion constant.  Here we
concentrate on the tri-interstitial finding an activation energy of
0.4-0.5 eV in DFT (and 0.6 eV in TB).
  
{\it Which tri-interstitial to study?}  In a previous work from our
research group~\cite{richie04}, the three lowest-energy
tri-interstitial geometries, denoted $I_{3}^{a}$, $I_{3}^{b}$, and
$I_{3}^{c}$, were identified by analysis of tight-binding molecular
dynamics (MD) simulations followed by density functional relaxations.
The TB calculations use the Lenosky \textit{et
  al.}~\cite{lenosky97tbpot} potential.  $I_{3}^{a}$ is the ground
state and $I_{3}^{b}$ and $I_{3}^{c}$ are excited states \footnote{It
  should be noted that our notation differs from that of Estreicher
  \textit{et al.}~\cite{estreicher01prl}; our $I_{3}^{b}$ geometry
  corresponds to their $I_{3}^{a}$.}; the density functional energies
for the three defects in 216+3 atom supercells were 2.24 eV/atom, 2.37
eV/atom, and 2.47 eV/atom respectively.  The tight-binding MD
simulations show only $I_{3}^{b}$ is a rapid diffuser, while
$I_{3}^{a}$ and $I_{3}^{c}$ can be formed by interconversion of
$I_{3}^{b}$, but are themselves immobile within the 5~ns time scale of
the simulations~\cite{richie04}.  Hence, we focus on elucidating the
microscopic diffusion process of the compact tri-interstitial
$I_{3}^{b}$.  $I_{3}^{b}$ has a compact defect geometry in which a
perfect tetrahedron of four atoms replaces a single atom in the
silicon lattice, with the faces oriented along the four symmetry
related $\langle 111 \rangle$ directions.

Tight-binding molecular dynamics determines an initial $I_{3}^{b}$
diffusion path.  Density-functional theory uses the
nudged-elastic-band method to refine the diffusion path and determines
the accurate diffusion barrier.  The self-diffusion of $I_{3}^{b}$ can
be visualized in terms of the atoms most distorted during the process.
In particular a five-atom object translates while rotating to avoid
adjacent atoms.  NEB finds four equivalent paths along $\langle 111
\rangle$ with a diffusion constant of $D = 4 \times 10^{-5}~\exp
(-0.49~{\rm eV} / k_{B} T)~{\rm cm^2/s}$.

\begin{figure}
\includegraphics[width=8cm]{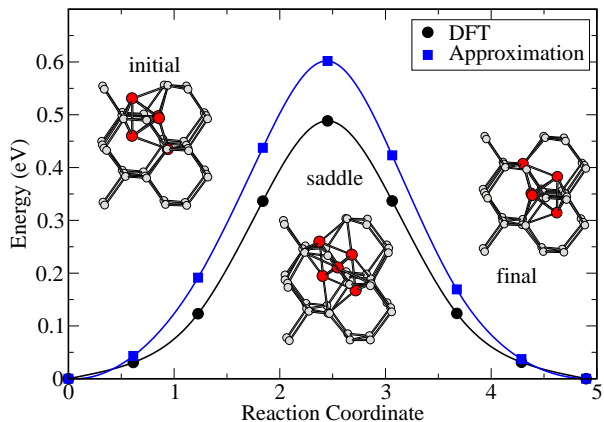}
\caption{\label{fig_diffpath} 
Transition paths of $I_{3}^{b}$ in  DFT-GGA and the 
approximation of collective motion of five atoms within DFT-GGA.
Five atoms are highlighted in gray for the initial state,
saddle point  and  final state in the insets, respectively.
}
\end{figure}

\textit{Minimum energy path / nudged elastic band.}
Figure~\ref{figure_I3b} shows two $I_3^b$ defects in neighboring sites
that correspond to initial and final configurations along the
diffusion path, the [111] direction.  We obtain the initial pathway
from analysis of molecular dynamics trajectories~\cite{richie04}.  The
climbing-image NEB (CI-NEB) method~\cite{cineb} refines an accurate
pathway and finds the diffusion barrier between these two $I_3^b$
minima.  The CI-NEB scheme guarantees that the image with the highest
energy converges to the saddle point.  All NEB calculations are
performed with the VASP density functional code~\cite{vasp1, vasp2}
employing the Perdew-Wang GGA functional~\cite{perdew}, and ultra-soft
Vanderbilt-type pseudopotentials~\cite{pseudo1} as provided by G.
Kresse and J.  Hafner~\cite{pseudo2}.  An initial relaxation of the
two end points initiates a full relaxation of seven images along the
path, keeping the volume of the cell fixed.  Energy and atomic
position convergence of 3~meV and 0.005~\AA, respectively, is
confirmed for 64+3 atom supercell by comparing the results for a
250~eV energy cutoff and a $3\times3\times3$ k-point mesh with a
300~eV cutoff and a $4\times4\times4$ mesh.  The seven images are
initialized by linear interpolation between the two relaxed end
points.  Each of the images is relaxed until the atomic and spring
forces are less than 10~meV/\AA.

Figure~\ref{fig_diffpath} shows the diffusion path with an activation
energy of 0.49~eV.  Harmonic transition state theory~\cite{vineyard}
calculates the defect jump rate $\Gamma$ from the phonon frequencies
at the minimum ${\nu_{i}^{min}}$ and at the saddle point
${\nu_{i}^{sad}}$ and the activation energy $\Delta E$
\begin{equation}
\Gamma = \Gamma_{0} ~e^{- \Delta E / k_{B} T},
\label{eq:vineyard}
\end{equation}
where the prefactor $\Gamma_{0}$ is given by: 
\begin{equation}   
\Gamma_{0}  =  \prod_{i}^{3N-3}~\nu_{i}^{min} / \prod_{i}^{3N-4}~\nu_{i}^{sad}. 
\label{eq:gamma-0}
\end{equation}
The dynamical matrix method within {\it ab initio} GGA density
functional theory determines the phonon frequencies.  Each atom is
displaced in the $\hat{{\rm x}}$, $\hat{{\rm y}}$, $\hat{{\rm z}}$
direction by 0.03~\AA\ and the calculated forces are used to construct
the Hessian matrix for the system, which is diagonalized to find the
phonon frequencies.  This yields a jump rate $\Gamma = 0.2~{\rm THz}
~\exp(- 0.49 / k_{B} T )$.  Calculations for a larger cell of $216+3$
atoms provide a diffusion barrier of 0.43~eV estimating a finite size
error of about 0.1~eV.

\begin{figure}
\includegraphics[width=1.5in]{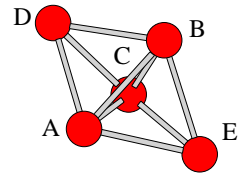}
\caption{\label{fig_five} 
  Five most displaced atoms during the diffusion.  These five atoms
  move collectively.  Atom D and E define an axis of rotation ([111]),
  around which atom A, B and C are rotating.}
\end{figure}

\begin{figure}
\includegraphics[width=8cm]{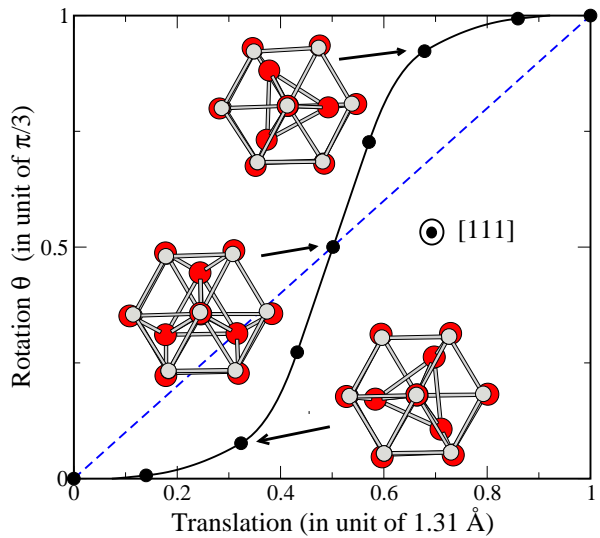}
\caption{\label{fig_2Dpath} 
  Transition path with seven images projected onto two-dimensional
  reduced coordinates.  The inset shows images 3, 5 (saddle point),
  and 7 along the transition path.}
\end{figure}

\begin{table}
  \caption{\label{tab_2Dpath}
    The translation and rotation of five atoms along the [111] 
    direction  with respect to the initial state for all seven images in 
    Fig~\ref{fig_2Dpath}.}
  \begin{tabular}{ccc}
    \hline
    \hline
    image   & Translation (\AA)  &  Rotation (degree)    \\
    \hline
    1 &  0.18 &   0.4 \\
    2 &  0.42 &   4.6 \\
    3 &  0.57 &  16.4 \\
    4 &  0.66 &  30.0 \\
    5 &  0.75 &  43.6 \\
    6 &  0.89 &  55.4 \\
    7 &  1.13 &  59.6 \\
    \hline
  \end{tabular}
\end{table}

\textit{Collective motion of atoms.} In the diffusion event, the five
most-displaced atoms move collectively, with a screw-like motion.  The
insets in Figure~\ref{fig_diffpath}, showing the atomic configurations
for initial, saddle and final states, indicate significant displacement
of the five solid-colored atoms.  All other atoms relax only slightly
during the diffusion with a maximum displacement of 0.18~\AA.

In order to illustrate the screw-like collective motion, we highlight
the five highly-displaced atoms in Figure~\ref{fig_five}.  Atom A, B,
and C are located in a $(111)$ plane, forming an equilateral triangle,
and translate along and rotate about the $[111]$ axis defined by atom
D and E.  Meanwhile, atoms D and E translate along the same $[111]$
direction.  Atoms D, A and B form an equilateral triangle, whose
normal vector indicates another diffusion direction, and all five
atoms form a double-tetrahedron.  The bond between A and B is
2.49~\AA\ at the minimum, 6\% longer than the perfect silicon-silicon
bulk bond distance.  During the diffusion event this bond length
varies by less than 4.5\%.  During defect migration the
double-tetrahedron translates 1.31~\AA\ and rotates 60 degrees as
shown in Figure~\ref{figure_I3b} and the insets of
Figure~\ref{fig_diffpath}.  Table~\ref{tab_2Dpath} lists the values
for the translation and rotation of the seven relaxed NEB images.

The diffusion event is approximated by a uniform translation and
rotation of a fixed double-tetrahedron.  For instance, rotating the
double-tetrahedron of the initial configuration (image 1) by 16.4
degrees, and translating it by 0.57~\AA~while keeping its geometry
fixed and relaxing the other atoms provides an approximation to image
3.  Applying this procedure to each of the images along the diffusion
path, we obtain the approximate diffusion path in
Figure~\ref{fig_diffpath}.  The approximate activation energy is
0.6~eV, only 20\% higher than the fully relaxed value.

Figure \ref{fig_2Dpath} shows the translation and rotation of the
double-tetrahedron during the diffusion.  Translation occurs before
the structure rotates. The double-tetrahedron is displaced half way
and rotated 30 degrees at the saddle point.  The insets show images 3,
5 (saddle point) and 7 viewed from the $[111]$ direction and illustrate
that three-fold symmetry persists during the diffusion.

\textit{Diffusion rate.} The defect $I_{3}^{b}$ can move to four
neighboring sites along the four symmetry-equivalent $\langle 111
\rangle$ directions.  The diffusion constant for this random walk is
$D = 2~\frac{4}{6} a^{2} \Gamma$ (derivation is given in
Ref.~\cite{diffconst}), where the jump rate $\Gamma$ is defined by
Eq.~\ref{eq:vineyard} and the displacement during the diffusion event
is $a = 1.31$~\AA.  The factor two takes into account the fact that
the defect can rotate either in clockwise or counter-clockwise
direction during the diffusion.  The resulting diffusion constant is
$D = 4 \times 10^{-5}~\exp (- 0.49~{\rm eV} / k_{B} T)~ {\rm cm^2/s} $.

\begin{table}
  \caption{\label{tab_sdt}
    Formation and activation energy of the neutral
    single-interstitial diffusing along the {\it X-T-X} path,
    and of the neutral ground state di-interstitial and
    compact tri-interstitial during self-diffusion.
    The formation energies are calculated in a $216+n$ atom supercell
    and the activation energies are calculated using NEB with
    a $64+n$ atom supercell.
    Details of the diffusion paths of single- and di-interstitials will
    be published elsewhere~\cite{pending}.}
  \begin{tabular}{ccc}
    \hline
    \hline
    & Formation energy /atom (eV)  &  Activation energy  (eV)  \\
    \hline
    $I_{1}$ &  3.810    &  0.29,  0.28 \footnote{NEB result from Ref.~\cite{lopez04} is calculated using 
      $216+1$ atom 
      supercell.}     
    \\
    $I_{2}$ &  2.827   &   0.30    \\
    $I_{3}^{b}$  &  2.368   &    0.49       \\
    \hline
  \end{tabular}
\end{table}

\textit{Discussion.} Table \ref{tab_sdt} summarizes the {\it ab
  initio} density functional theory results for formation and
activation energy of $I_{1}$, $I_{2}$, and compact $I_{3}$ defects.
The theoretical values of 0.28 to 0.49 eV are considerably lower than
the experimental results from metal-diffusion experiments which find
activation energies, believed to be for single interstitials,
typically ranging from $\sim$1.4 to 1.8 eV \cite{bracht99cep}.
However, a recent experiment~\cite{partyka} identifies enhanced
diffusion of single interstitials at 150~K, while noting it is
inconsistent with results from earlier approaches.  This lower
temperature would be consistent with the NEB result for the energy
barrier of single interstitial diffusion of 0.3~eV.

Our pathway for $I_3$ diffusion is the same as that found by Cogoni
{\it et al.}~\cite{cogoniprb05} in the $64+3$ atom supercell.  On the
other hand, our DFT barriers for $I_1$, $I_2$, and $I_3$ of 0.3, 0.3
and 0.4-0.5 eV are much smaller than the tight-binding results by
Cogoni {\it et al.}~\cite{cogoniprb05} of 0.9, 0.9 and 1.7 eV.  We
also perform the NEB calculation to examine the diffusion path of the
compact tri-interstitial within Kwon's potential~\cite{kwon}, and
obtain a diffusion barrier of 1.2 eV in contrast to 0.6 eV barrier
within Lenosky's TB potential~\cite{lenosky97tbpot}. The Kwon's
potential overestimates the phonon frequencies~\cite{lenosky97tbpot},
which makes a local minimum steeper. This suggests that Kwon's
potential will consequently overestimate the barrier for the interstitial
diffusion, which is characterized by the small displacements of
several atoms deviating from the equilibrium sites.

We perform density of states (DOS) calculations for the interstitial
defect along the diffusion path within a $216+3$ atom supercell.  The
DOS is calculated for the minimum, the saddle point and an
intermediate structure.  The DOS shows a band gap of 0.71-0.74 eV for
all three configurations along the diffusion pathway.  There are no
states in the gap.  The size of the gap is close to the band gap of
pure Si in GGA of 0.72 eV.  The lack of gap states indicates that
charge transfer may not play a significant role for the diffusion
of a compact tri-interstitial.

While the formation energy of $I_{3}$ is lower than $I_{1}$ on a
per-atom basis, it is much higher when considered on a total basis.
One implication that can be drawn from this is that under equilibrium
conditions, $I_{3}$ is not present in any significant quantity, but
under conditions which inject excess
interstitials~\cite{eaglesham94apl}, such as ion implantation, it may
be present in significant amounts. Our results have possible relevance
for modeling transient enhanced diffusion which follows ion
implantation~\cite{martin}.

\textit{Conclusions.}  We have elucidated the pathway for the
diffusion of the compact tri-interstitial, $I_{3}^{b}$, the only
fast-diffusing tri-interstitial species in crystalline silicon.
During the self-diffusion event, five atoms move collectively in a
screw-like motion along one of four symmetry-related $\langle 111
\rangle$ directions.  Our DFT result shows a low activation energy of
0.49 eV and a diffusion constant of $ 4 \times 10^{-5}~\exp (-
0.49~{\rm eV} / k_{B} T)~ {\rm cm^2/s}$.  Under conditions such as ion
implantation that creates excess interstitials and hence favor cluster
formation, $I_{3}^{b}$ diffusion may be an important process due to
the low activation energy.

\textit{Acknowledgments.}  The work was supported in part by DOE-Basic
Energy Sciences, Division of Materials Sciences (DE-FG02-99ER45795).
Computing resources were provided by the Ohio Supercomputing Center.
We also wish to acknowledge helpful conversations with Jeongnim Kim.

\end{document}